\documentclass[12pt]{iopart}
\usepackage{graphicx,graphics,color,epsfig}
\usepackage{iopams}

\begin{document}

\title[Pseudospin and spin-spin interactions in ultra-cold alkali atoms]{Pseudospin
and spin-spin interactions in ultra-cold alkali atoms}

\author{D. H. Santamore\dag\ and Eddy Timmermans\ddag}

\address{\dag\ Department of Physics, Temple University, Philadelphia, PA 19122, USA}

\address{\ddag\ CNLS, Los Alamos National Laboratory, Los Alamos, NM 87545, USA}

\begin{abstract}
Ultra-cold alkali atoms trapped in two distinct hyperfine states in
an external magnetic field can mimic magnetic systems of spin $1/2$
particles.  We describe the spin-dependent effective interaction
as a spin-spin interaction.  As a consequence of the zero-range,
the interaction of spin $1/2$ bosons can be described as an Ising or,
alternatively, as an XY-coupling.  We calculated the spin-spin
interaction parameters as
a function of the external magnetic field in the Degenerate Internal
State (DIS) approximation.  We illustrate the advantage of the spin-spin
interaction form by mapping the system of N spin $1/2$ bosons confined
by a tight trapping potential on that of N spin $1/2$ spins coupled via an infinite
range interaction.

\end{abstract}


\maketitle

\section{Introduction}

The ability to mimic quantum spins and the promise for controlling their
mutual interactions are central ingredients of the cold atom prospects for
simulating complex quantum systems: Quantum computation schemes rely on the
entanglement of qubits \cite{Chuang}, which act as spin $\frac{1}{2}$ objects.
Lattice spin models have provided successful descriptions of strongly correlated
electron systems \cite{Sachdevpaper}, usually with the parameters
situated near a quantum phase transition so that the systems exhibits
quantum critical behavior. A general description of quantum phase transitions
has not been devised yet \cite{Sachdevbook} and while scaling arguments have
been applied \cite{quantumscaling}, a general renormalization scheme remains
an outstanding challenge. The experimental inaccessibility of the condensed
strongly correlated electron systems certainly contributes to the challenge
of understanding quantum phase transition physics. Cold atom experiments,
which offer a very different accessibility and a very different set of control knobs
may offer a different and independent laboratory to test concepts.
Spin $\frac{1}{2}$ degrees of freedom are also essential building blocks of quantum
fluids in regimes that are novel for table-top experiments such as
Cooper-paired fermion fluids in the BEC-BCS crossover regime
\cite{MCSLSSK99,PSKJH05,PWLH07}, and polarized fermion superfluids (for
example, see theory: \cite{AGIM02,EV02,XGLW03,A06,KMJT06,
PW07,MOYDOM08,KSSS09,BDNJ09}\ and experiments: \cite{SSSK08,SSSK208,PWLHHS06,
LPLH09,BARJCDG04,WRAKSDG07}.) In addition, bosons trapped in two
hyperfine states can act as spin $\frac{1}{2}$ particles \cite{Legg}.  In Bose-Einstein
condensates (BEC's) of such particles, the formation of spin $\frac{1}{2}$
magnetic domains has been observed \cite{Ketterle1}, which is a phase
separation transition \cite{Eddy}.

The observation of spin domain formation \cite{Ketterle1}, of intricate
multi-component spin physics in BECs \cite{Wiemel1}, of unstable spin dynamics
in spin $1$ BECs \cite{Stamper} and the engineering of two-component BEC
vortex states \cite{Wiemel2} and \cite{Wiemel22} highlight the cold atom
ability to access, populate and control hyperfine spin states in cold atom
traps. Other experiments demonstrated the cold atom ability to select specific
hyperfine spin states in a static, external magnetic field as effective spin
components: The above mentioned observation of BEC-phase separation, the
probing of the BCS-BEC crossover regime, the spin polarized fermion
superfluids, and the recently observed phase transition in an itinerant
ferromagnet-like system were all carried out with effective spin $\frac{1}{2}$
particles that occupy a linear combination of hyperfine spin eigenstates in
the presence of a static external magnetic field. For reasons we describe
below, we refer to the effective spin as `pseudospin'.

In this paper, we derive the short-range, effective pseudospin-spin interaction
potential that describes $s$-wave interactions of ultra-cold atoms that occupy a
superposition of two hyperfine states in an external magnetic field.  The resulting
spin-spin interaction has an anisotropic form, but the commutation or anticommutation
relations that respectively characterize bosonic and fermionic systems, combined with
the short-range (delta-function) nature of the interaction project the interactions
onto the pseudospin triplet subspace for bosons and on the pseudospin
singlet subspace for fermions.   For bosonic atoms, we find that the inter-particle
interactions can be described by three terms: the first term is a short-range spin
independent interaction potential; the second term describes the coupling of
one particle's pseudospin to a short-range, spin-independent effective
magnetic field carried by the other particle; and the third term is a short-range
Ising spin-spin coupling.  Alternatively, the spin-spin interaction can be cast into
the form of an $XY$-coupling.  We show how these terms can be calculated
in the Degenerate Internal State (DIS) approximation \cite{Boud1}.  We illustrate the advantage
of the spin-spin interaction description by deriving the many-spin hamiltonian of N
boson particles contained in a tightly confining trap.  This system, a controllable quantum
magnet, is a promising system to probe macroscopic quantum tunneling, realize
spin squeezing and Heisenberg-limited interferometry \cite{Dowl}.

\section{Cold atom spins}

The first question to be addressed in simulating one system by another is:
what are the salient features of the simulated system that the simulating system
needs to possess?  As we see below, a spin operator has three components
that satisfy a particular commutator algebra.  Any operator that satisfies this
algebra is a candidate to acts as a spin.  We construct the available alkali
atom spin states in an external magnetic field explicitly.  By selecting two of these states
the experimentalists create a cold atom system of effective spin $\frac{1}{2}$ particles.

\subsection{General spin properties: spin and rotation}

We denote the
quantum state of a spin $\overrightarrow{\xi}$ by the spinor $\left\vert \xi\right\rangle $.
A spin of amplitude $\xi$, where $\xi$ is an integer or half-integer number, has $2\xi+1$
eigenstates  $\left\vert \xi_{j} \right. \rangle $ of the spin projection
operator $\hat{\xi}_{z} \left\vert \xi_{j} \rangle \right. = j \left\vert \xi_{j} \rangle \right.$ with $j=-\xi,-\xi+1,...,\xi-1,\xi$.
The spin operator is the generator of rotations so that
under a rotation $\mathcal{R}_{\vec{\epsilon},\alpha}$ by an angle $\alpha$
around the direction of unit vector $\vec{\epsilon}$, the spin state  $\left\vert \xi\right\rangle $
(which is a superposition of the states $\left\vert \xi_{j} \rangle \right.$) transforms according to
\begin{equation}
\left\vert \mathcal{R}_{\vec{\epsilon},\alpha}\left(  \xi\right)
\right\rangle =\exp\left(  i\alpha\vec{\epsilon}\cdot\overrightarrow{\xi
}\right)  \left\vert \xi\right\rangle.
\label{rot}
\end{equation}
For an infinitesimal rotation, $\alpha = \delta \alpha \ll 1$, to lowest order
in $\delta \alpha$,
\begin{equation}
\left\vert \mathcal{R}_{\vec{\epsilon},\delta \alpha}\left(  \xi\right)
\right\rangle \approx \left[ 1 + i\delta\alpha\vec{\epsilon}\cdot\overrightarrow{\xi
} \right] \left\vert \xi\right\rangle.
\label{infrot}
\end{equation}
Quantum mechanically,  an operator $\mathcal{O}$ in the Hilbert space
of a single spin is characterized by its matrix elements
$\left\langle \xi_{i}|\mathcal{O}|\xi_{j}\right\rangle $ where $\left\vert
\xi_{i}\right\rangle $ is the $i$-th element of the $\left\{  \xi_{\xi}%
,\xi_{\xi-1},...\right\}  $, basis. We relate the basis matrix elements in
the original and in the rotated spinor
basis $|\xi_{i}^{\prime}\rangle$. When the rotated frame
is obtained by rotating around $\vec{\epsilon}$ over an angle $-\delta \alpha$,
\begin{equation}
\langle\xi_{i}^{\prime}|\mathcal{O}|\xi_{j}^{\prime}\rangle\approx\langle
\xi_{i}|\left(  \mathcal{O}-i\;\delta\alpha\;\left[  \mathcal{O}%
,\overrightarrow{\epsilon}\cdot\overrightarrow{\xi}\right]  _{-}\right)
|\xi_{j}\rangle\;,
\end{equation}
where the subscript ($-$) outside the bracket indicates the usual commutator.
As a consequence, the operator $\mathcal{O}$ transforms as
\begin{equation}
\mathcal{O}^{\prime}=\mathcal{O}-i\;\delta\alpha\;\left[  \mathcal{O}
,\overrightarrow{\epsilon}\cdot\overrightarrow{\xi}\right]  _{-}.
\label{rotop}%
\end{equation}
Moreover, if the $\mathcal{O}$--operator is rotationally invariant,
$\mathcal{O}^{\prime}=\mathcal{O}$, then $\left[  \mathcal{O},\overrightarrow
{\epsilon}\cdot\overrightarrow{\xi}\right]  _{-}=0$ is the necessary and
sufficient condition to ensure that the $\mathcal{O}$--operator is invariant
under a rotation around the $\overrightarrow{\epsilon}$-vector.

Classically an infinitesimal rotation transforms a general
vector $\overrightarrow{v}$ to $\mathcal{R}_{\overrightarrow{\epsilon},\delta\alpha}\left(  \overrightarrow
{v}\right)  =\overrightarrow{v}+\;\delta\alpha\;\overrightarrow{\epsilon
}\times\overrightarrow{v}$.  If spin is to act as a vector under a rotation,
it is to transform as
\begin{equation}
\mathcal{R}_{\overrightarrow{\epsilon},\delta\alpha}\left(  \overrightarrow
{\xi}\right)  =\overrightarrow{\xi}+\;\delta\alpha\;\overrightarrow{\epsilon
}\times\overrightarrow{\xi}. \label{rotvec}
\end{equation}
Choosing
$\mathcal{O}=\overrightarrow{\xi}$, by combining Eqs.\ (\ref{rotop}) and
(\ref{rotvec}), and identifying the terms linear in $\delta\alpha$ we obtain
\begin{equation}
i\;\overrightarrow{\epsilon}\;\times\;\overrightarrow{\xi}=\left[
\overrightarrow{\xi},\overrightarrow{\epsilon}\cdot\overrightarrow{\xi
}\right]  _{-}.
\end{equation}
The scalar product of the above vector equality with a unit vector
$\overrightarrow{\eta}$ gives
\begin{equation}
\left[  \overrightarrow{\xi}\cdot\overrightarrow{\eta},\overrightarrow{\xi
}\cdot\overrightarrow{\epsilon}\right]  _{-}=i\;\overrightarrow{\xi}
\cdot\left(  \overrightarrow{\eta}\times\overrightarrow{\epsilon}\right),
\end{equation}
and choosing the $\vec{\eta}$ and $\vec{\epsilon}$ vectors as part
of a Cartesian $(x,y,z)$ reference frame, $\vec{\eta}=\vec{x}$ and $\vec{\epsilon}=\vec{y}$, we obtain
\begin{equation}
\left[  \hat{\xi}_{x},\hat{\xi}_{y} \right] _{-}=i\;\hat{\xi}_{z} ,
\label{alg}
\end{equation}
the usual form of the angular momentum commutator relation.
From this commutator, it can be shown that $\hat{\xi}_{x}\pm\hat{\xi}_{y}$ are
raising and lowering operators, increasing and decreasing the spin projection
eigenvalue of $\hat{\xi}_{z}$ by one unit.  As operators evolve according
to the Heisenberg commutator equations in the quantum evolution of the
system, any operator satisfying the above commutator relations will
evolve as a spin (or angular momentum) of the same Hamiltonian.

The pseudospins we describe below are \textit{not} the generators of rotation
and their mutual interactions are \textit{not} rotationally invariant.
However, the pseudospins can be described by pseudospin operators that
\textit{do} satisfy an angular momentum algebra. In the Heisenberg picture,
the quantum equations of motion shows that the dynamics is identical to that
of real spins except that the spins and their mutual interactions are
\textit{not} rotationally invariant.

\subsection{The spin structure of alkali atoms in a magnetic
field\label{Sec_spinstructure}}

Alkali atoms have two spin variables: the electron spin $\mathbf{s}_{e}$ of
magnitude $s_{e}=\frac{1}{2}$ carried by the $s$-wave valence electron and the
nuclear spin $\mathbf{i}$ of magnitude $i$. The electron 'spins'
in the magnetic field of the nucleus and the short-range part of the
nuclear spin's magnetic field (the Fermi contact term) contributes to the energy
in a first-order perturbation calculation \cite{Foote}. The corresponding
contribution takes the form $a_{hf}\mathbf{i}\cdot\mathbf{s}_{e}$, which is
the hyperfine interaction. We refer to $a_{hf}$ as the hyperfine energy.

In a static, external, homogeneous magnetic field $\mathbf{B}$ of strength $B
$, $\mathbf{B}=B\mathbf{z}$, the interaction of the electronic and nuclear
spins with the magnetic field and with each other are described by the Zeeman
spin Hamiltonian
\begin{equation}
\mathcal{H}=B\mathbf{z}\cdot\left[  \mu_{e}\mathbf{s}_{e}-\mu_{N}
\mathbf{i}\right]  +a_{hf}\mathbf{s}_{e}\cdot\mathbf{i},
\label{Zeeman1}
\end{equation}
where $\mu_{e}=\hbar/2m_{e}c$, $\mu_{N}=\hbar/2m_{p}c$ represent the
electronic and nuclear Bohr magneton, respectively, with $m_{e}$ for the electron mass
and $m_{p}$ for the proton mass. As
the nucleons are heavier and the nuclear Bohr magneton, $\mu_{N}$,
proportionally smaller (by three orders of magnitude) than $\mu_{e}$, $\mu
_{N}$ and we neglect it for now.  Scaling the magnetic field strength in units of the
hyperfine field $B_{hf}=a_{hf}/\mu_{e}$, $b=B/B_{hf}$, the Zeeman
Hamiltonian takes the form
\begin{equation}
\frac{\mathcal{H}}{a_{hf}}=\mathbf{s}_{e}\cdot\mathbf{i}+b\;\mathbf{z}%
\cdot\mathbf{s}_{e}.
\label{zeeman}
\end{equation}
At zero magnetic field, $b=0$, the structure of the spin Hamiltonian
eigenstates is most readily analyzed by introducing the total hyperfine spin
operator, $\mathbf{f}=\mathbf{i}+\mathbf{s}_{e}$. The Hamiltonian eigenstates
can be chosen to be eigenstates of good $\left(  \mathbf{f}^{2}\right)  $ and
$\mathbf{f}\cdot\mathbf{z}$ quantum numbers. The eigenvalues of the
$\mathbf{f}^{2}$ operator are $f \left( f + 1 \right)$, with $f=i\pm1/2$. The eigenvalues of the
$\mathbf{f}\cdot\mathbf{z}$ are labeled by the eigenvalue $m_{f}$. The $|f,m_{f}\rangle$ states
of the same $f$-quantum number are degenerate. The first term of Eq.\ (\ref{zeeman}) can be
written as $\left(  \mathbf{f}^{2}-\mathbf{s}_{e}^{2}-\mathbf{i}^{2}\right)
/2$, and when $b=0$, $E(f,m_{f},b)=E_{0}(f)$ and $E_{0}(f=i+1/2)=a_{hf}
i/2$ and $E_{0}(f=i-1/2)=-a_{hf}(i+1)/2$, giving a hyperfine splitting
$a_{hf}(i+\frac{1}{2})$.  This splitting is determined spectroscopically. For example, $^{23}Na$
has a hyperfine energy measured to be $a_{hf}=42.5mK=0.95GHz$ (which greatly exceeds
cold atom trap depths $\sim\mu K$), corresponding to a hyperfine magnetic field strength
equal to $B_{hf}\approx 709 G$.

At finite magnetic field, $f$ is not a good quantum
number, but it is convenient and customary to refer to the $f$-value that the
same state takes on in the adiabatic $b\rightarrow 0$-limit. On the other
hand, the total spin projection, $m_{f}$ remains a good quantum number. We
parametrize the corresponding state by an angle $\theta$ as
\begin{eqnarray}
\left\vert f^{\pm },m_{f}\right\rangle  &=&\pm \cos \left( \frac{\theta
^{\pm }}{2}\right) \left\vert m_{i}=m_{f}-\frac{1}{2},m_{e}=+\frac{1}{2}%
\right\rangle   \nonumber\\
&&+\sin \left( \frac{\theta ^{\pm }}{2}\right) \left\vert m_{i}=m_{f}+\frac{1%
}{2},m_{e}=-\frac{1}{2}\right\rangle.
\label{parame}
\end{eqnarray}
We restrict the range of $\theta$-angles to the $\left[  0,\pi\right]  $ interval,
allowing for coefficients of a different sign by adding $\pm$ explicitly to the cosine term.
By parametrizing the state as in Eq.\ (\ref{parame}), the $\theta$ is also the inclination
angle of the average electron spin vector in that hyperfine state.
The optimal value of $\theta$ depends on the spin projection eigenvalue $m_{f}$.

Using the angular momentum algebra Eq.\ (\ref{alg}), the spin product expectation value
$\langle\mathbf{s}_{e}\cdot\mathbf{i}\rangle$, of the states of
Eq.\ (\ref{parame}) give
\begin{equation}
\left\langle \mathbf{s}_{e}\cdot\mathbf{i}\right\rangle =\frac{1}{2}\left[
\cos\left(  \theta\right)  m_{f}\pm\sin\left(  \theta\right)  \sqrt{\left(
i+\frac{1}{2}\right)  ^{2}-m_{f}^{2}}\right]  -\frac{1}{4},
\end{equation}
where, $\pm$ refer to the relative sign of the electron spin up and down
components, as introduced in (\ref{parame}). Taking the extreme of the expectation value of Eq.\ (\ref{zeeman})
with respect to $\theta$ in (\ref{parame}) gives
\begin{equation}
\frac{E(\pm,m_{f})}{a_{hf}}=\left(  b+m_{f}\right)  \cos\left(  \theta\right)
\pm\sin\left(\theta\right)\sqrt{\left(  i+\frac{1}{2}\right)  ^{2}-m_{f}^{2}}-\frac{1}{4},
\end{equation}
and by requiring $\partial E/\partial\theta=0$, we obtain the optimal inclination angle $\tan\left(
\theta^{\pm}\right)  $,
\begin{equation}
\tan\left(  \theta^{\pm}\right)  =\pm\frac{\sqrt{\left(  i+\frac{1}{2}\right)
^{2}-m_{f}^{2}}}{\left(  b+m_{f}\right)  }.
\end{equation}
As $b\gg1$, $\tan\left(  \theta^{\pm}\right)  \rightarrow\pm\lbrack
i+1]/b\rightarrow\pm\infty$, corresponding to $\theta^{+}\rightarrow0$ and
$\theta^{-}\rightarrow\pi$: the electron spin in hyperfine states align or
anti-align with the external magnetic field when the strength of that field
significantly exceeds the hyperfine field strength. For notational convenience
we introduce the square root of the sum of the squares of the numerator and
denominator,
\begin{equation}
e\left(  m_{f},b\right)  =\sqrt{b^{2}+2m_{f}b+\left(  i+\frac{1}{2}\right)
^{2}}.
\end{equation}
Then the angular projections can be written as
\begin{equation}
\cos\left(  \theta^{\pm}\right)  = \pm \frac{\left(  b+m_{f}\right)  }{e\left(
m_{f},b\right)  },\nonumber\\ \; \; \; \; \; \;
\sin\left(  \theta^{\pm}\right)  =  \frac{\sqrt{\left(  i+1/2\right)
^{2}-\left(  m_{f}\right)  ^{2}}}{e\left(  m_{f},b\right)  }.
\label{cossin}
\end{equation}
Substitution of Eq.\ (\ref{cossin}) into Eq.\ (\ref{zeeman}) yields the Zeeman
eigenenergies $E\left(  \pm,m_{f},b\right)$ ,
\begin{equation}
\frac{E\left(  \pm,m_{f},b\right)  }{a_{hf}}=\pm\frac{e\left(  m_{f},b\right)
}{2}-1/4,
\label{Zeeman2}
\end{equation}
where the superscripts $\pm$ are now seen to indicate whether the hyperfine
state belongs to the $f^{+}=i+1/2$ or the $f^{-}=i-1/2$--superpositions of Eq.\ (\ref{parame}).
To see that, note that in the limit of vanishing magnetic field, $\lim
_{b\rightarrow0}$, $e(m_{f},b)\rightarrow i+1/2$, so that $E(+,m_{f},b)$
approaches $E_{0}\left(  f=i+1/2\right)  =a_{hf}i/2$ while $E(-,m_{f},b)$
approaches $E_{0}\left(  f=i-1/2\right)  =-a_{hf}\left[  i+1\right]  /2$.
Actually, the expression Eq.\ (\ref{Zeeman2}
) describes the `non-stretched' states, $\left\vert m_{f}\right\vert <i+1/2$.
For the stretched states $\left\vert f=i+1/2,m_{f}=i+1/2\right\rangle $ and
$\left\vert f=i+1/2,m_{f}=-i-1/2\right\rangle $,
\begin{equation}
\frac{E\left(  f^{+},m_{f}=\pm f^{+},b\right)  }{a_{hf}}=\pm\frac{b}{2}
+\frac{i}{2},
\end{equation}
corresponding to $\theta=0$ and $\theta=\pi$ in the linear superpositions of
Eq.\ (\ref{parame}).
\begin{figure}[ptb]
\begin{center}
\includegraphics[width=3in] {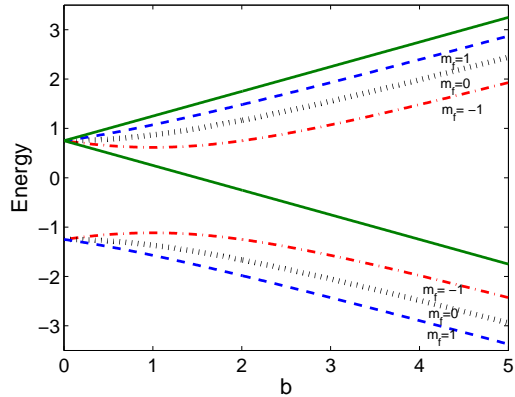}
\end{center}
\caption{Hyperfine energy levels for nuclear spin $i=3/2$, the nuclear spin
magnitude of bosonic atoms such as $^{7}Li$, $^{23}Na$ and $^{87}Rb$.}%
\label{Fig1boson}
\end{figure}
In Fig.\ \ref{Fig1boson}, we show the magnetic field dependence
of the spin states for the nuclear spin value $i=\frac{3}{2}$ that describes
$^{7}Li $, $^{23}Na$ and $^{87}Rb$, three of the most used cold atom alkali
atom species.

The electron spin expectation value $\langle\mathbf{s}_{e}\rangle$,
\begin{equation}
\left\langle f,m_{f}\left\vert \mathbf{s}_{e}\right\vert f,m_{f}\right\rangle
=\mathbf{z}\frac{\cos\left(  \theta\right)  }{2}=\pm\mathbf{z} \frac{\left(
b+m_{f}\right)  }{2e\left(  m_{f},b\right)  },
\label{inclin}
\end{equation}
points in the direction of the external magnetic field.

\subsection{Pseudospin\label{Sec_pseudospin}}

By trapping cold atoms in a specific hyperfine state in an external magnetic
field and by coherently transferring part of the spin population into a different
hyperfine state, either by accessing a laser-driven two-photon Raman transition or by means
of another magnetic field that is oscillating near the energy-difference, the
experimentalists prepare the atoms in a superposition of two hyperfine
states, $\left\vert f_{1},m_{f_{1}}\right\rangle $ and $\left\vert f_{2},m_{f_{2}}\right\rangle $.
If the hyperfine states have been selected to ensure that
no other spin states couple, $\left\vert f_{1},m_{f_{1}}\right\rangle $ and
$\left\vert f_{2},m_{f_{2}}\right\rangle $ act as the basis of an effective spin $\frac{1}{2}$.
We can than assign one the role of `spin-up' the other that of `spin-down' state,
$\left\vert f_{1},m_{f_{1}}\right\rangle =\left\vert \uparrow \rangle\right.$ and
$\left\vert f_{2},m_{f_{2}}\right\rangle = \left\vert \downarrow \rangle\right.$.
Interestingly, if the atoms are bosonic, i.e. if their hyperfine spin $f$ takes on an integer
value, the atoms become effective spin $\frac{1}{2}$ bosons.

The coupling to other hyperfine states has to be avoided because not only
does the role of additional
hyperfine states require a larger Hilbert space description, but also it generally causes
significant particle loss. For magnetic field strengths near the hyperfine
magnetic field, $b\sim1$, the Zeeman levels are widely spaced with energy
differences comparable to $a_{hf}$ (tens of $mK$), considerably larger than
the typical trap depths ($\mu K$). Hence, spin flip collisions in which
particles end up in lower energy spin states create particle pairs of
sufficient kinetic energy to evict spin flipped particles from the trap. However,
conservation of energy and conservation of total spin projection ($m_{f}$) in
binary atom interactions limit the spin-changing collisions that can take
place in an external magnetic field.

In the case of nuclear spin $i=\frac{3}{2}$, the experimentalist can select
either the two lowest or the two highest energy levels of the $f=1$ manifold:
either $\left\vert 1,0\right\rangle $ and $\left\vert 1,-1\right\rangle $ or
$\left\vert 1,0\right\rangle $ and $\left\vert 1,+1\right\rangle $.  While collisions
$\left\vert
1,0\right\rangle +\left\vert 1,0\right\rangle $ $\rightarrow\left\vert
1,-1\right\rangle +\left\vert 1,+1\right\rangle $ would produce spin states
not included in the spin-up and spin-down basis, in the region $b\sim1$,
the energy cost of a spin flip up ($m_{f}$ from $0$ to $+1$) outweighs the energy gain
from the corresponding spin flip down process ($m_{f}$ from $0$ to $-1$), as
can be seen from Fig.~\ref{Fig1boson} so that this process is energy forbidden.
For nuclear spin $1$, $i=1$, as in the case of $^{6}Li$, $^{23}$Na and $^{87}$Rb,
the experimentalist
can select the two lowest energy hyperfine states, $\left\vert \uparrow
\right\rangle =|\frac{1}{2},\frac{1}{2}\rangle$, and $\left\vert \downarrow
\right\rangle=\left\vert \frac{1}{2},-\frac{1}{2}\right\rangle $, for instance.

In the $\left\vert f_{1},m_{f_{1}}\right\rangle =\left\vert \uparrow \rangle\right.$ and
$\left\vert f_{2},m_{f_{2}}\right\rangle = \left\vert \downarrow \rangle\right.$ basis
the pseudospin operator takes the form $\mathbf{s}=\frac{1}
{2}\mathbf{\sigma}$, where $\mathbf{\sigma}$ is the Pauli-spin vector
operator, $\mathbf{\sigma}=\sigma_{x}\mathbf{x}+\sigma_{y}\mathbf{y}
+\sigma_{z}\mathbf{z}$, with $\sigma_{x}$, $\sigma_{y}$, and $\sigma_{z}$ the
Pauli spin matrices. Note that $\mathbf{s}$ is \textit{not} the generator of
rotations. Under rotation, the spin components do not transform among each
other according to Eq.~(\ref{rot}), although the $\mathbf{s}$ components still
satisfy angular momentum commutator relations $\left[  s_{x},s_{y}\right]
_{-}=is_{z}$. Therefore, the time evolution of the many-spin system (governed
by commutation relations in the Heisenberg picture) is indistinguishable from
the quantum evolution of the magnetic many-spin system of the same parameters.
The concept of effective spin parallels closely that of iso-spin in nuclear
physics \cite{DeBenedetti}, except that a Raman transition and/or oscillating
magnetic field can convert spin-down into spin-up particles. Likewise, as
emphasized by Bloch \cite{Bloch}, any two-state system (and its decoherence)
can be described by an effective spin $\frac{1}{2}$ system.

To picture the role of the pseudo-spin direction, we parametrize
an arbitrary, normalized two component spinor $\left\vert \xi_{ps}
\right\rangle $ by introducing two angles $\theta$ and $\phi$.
\begin{equation}
\left\vert \xi_{ps}\right\rangle =\left(
\begin{array}
[c]{l}
\xi_{\uparrow}\\
\xi_{\downarrow}
\end{array}
\right)  =\left(
\begin{array}
[c]{l}
e^{i\frac{\phi}{2}}\cos\frac{\theta}{2}\\
e^{-i\frac{\phi}{2}}\sin\frac{\theta}{2}
\end{array}
\right)  .
\end{equation}
In this notation, $\phi$ denotes the relative phase: the difference of the
complex phase of the `up'\ amplitude and of that of the `down'\ amplitude.
Also, $\cos\theta$ denotes the effective polarization: the difference between
the up probability and the down probability. Then the expectation value of the
effective spin vector $\left\langle \mathbf{s}\right\rangle $ in terms of
$\theta$ and $\phi$ is
\begin{equation}
\left\langle \mathbf{s}\right\rangle =\frac{1}{2}\left\langle \xi
_{ps}\right\vert \vec{\sigma}\left\vert \xi_{ps}\right\rangle =\frac{1}%
{2}\left(  \sin\theta\cos\phi\mathbf{x}+\sin\theta\sin\phi\mathbf{y}%
+\cos\theta\mathbf{z}\right)  ,\nonumber
\end{equation}
which is a vector on the surface of a sphere of radius $\frac{1}{2}$ (known as
the `Bloch sphere'). The inclination $\theta$ and azimuthal $\phi$ angles of
the expectation value of the pseudospin vector then respectively characterize
the `polarization' and the phase difference. If an ensemble measurement
reveals $N_{\uparrow(\downarrow)}$ atoms in the `up' (`down') hyperfine state,
then $\left\langle \cos(\theta)\right\rangle =\left(  N_{\uparrow
}-N_{\downarrow}\right)  /\left(  N_{\uparrow}+N_{\downarrow}\right) $.

\section{Spin-spin binary atom interactions}

Cold alkali atoms are interesting building blocks for simulating
magnetic systems: their mutual interactions preserve the overall spin
projection (the sum of $m_{f}$ is preserved), the interactions are
naturally spin dependent and the strength of these interactions and their
spin-dependence can be varied by a Feshbach resonance.  We write the effective atom-atom
interaction as a spin-spin interaction.  The statistics of indistinguishability,
combined with the short-range nature of the effective interaction gives
a short-range Ising-like spin-spin interaction or, alternatively, a short-range
$XY$-interaction.  We calculate the spin-dependence
in the Degenerate Internal State (DIS) approximation.

\subsection{Effective inter-particle interaction in pseudospin language}

In cold atom experiments, the length scales relevant to the
many-body physics description ($\sim\mu m$) significantly exceed the length
scale on which the atoms interact (which ranges from Bohr-radius to $nm$). As a
consequence, the interactions of indistinguishable atoms $i$ and $j$
with position coordinates $\mathbf{x}_{i}$ and $\mathbf{x}_{j}$ can be described by
an effective contact interaction potential
\begin{equation}
V\left(  \mathbf{x}_{i},\mathbf{x}_{j}\right)  =\frac{4\pi\hbar^{2}}{m}
a\delta\left(  \mathbf{x}_{i}-\mathbf{x}_{j}\right)  . \label{delta}
\end{equation}
where $a$ denotes the scattering length and $m$ represents the single particle
mass. In the low energy regime of interest in a many-body description, this interaction
reproduces the real binary atom scattering physics (to all orders) in a first order
perturbation calculation. The interaction potential reproduces the correct binary atom
s-wave scattering amplitude in the Born approximation.  Introducing the annihilation
(creation) field operators $\hat{\psi}$ ($\hat{\psi}^{\dagger}$), the particle-particle
interactions are described by
\begin{eqnarray}
\hat{H}_{int} &=& \frac{1}{2} \int d^{3} x \int d^{3} x' \; \;
\hat{\psi}^{\dagger}({\bf x}) \hat{\psi}^{\dagger}({\bf x}')
V\left({\bf x}-{\bf x}'\right) \hat{\psi}({\bf x}') \hat{\psi}({\bf x})
\nonumber \\
&=& \frac{1}{2} \left( \frac{4\pi \hbar^{2}}{m} \right) \int d^{3} x \; \;
\hat{\psi}^{\dagger}({\bf x}) \hat{\psi}^{\dagger}({\bf x})
\hat{\psi}({\bf x}) \hat{\psi}({\bf x}) \; ,
\end{eqnarray}
in the Hamiltonian operator.
When the indistinguishable particles are bosons that occupy two possible spin states,
$\left\vert\uparrow\right\rangle$ and $\left\vert\downarrow\right\rangle$, we distinguish interactions between
particles in like spin states, described by a scattering length $a_{\uparrow}$ ($a_{\downarrow}$)
if that state is the `up' (`down') spin-state, and between particles in unlike spin states,
described by scattering length $a_{u}$.  The Hamiltonian operator that accounts for these
interactions takes the form
\begin{eqnarray}
\hat{H}_{int}  & =\frac{1}{2}\left(  \frac{4\pi\hbar^{2}}{m}\right)  \left(
a_{\uparrow}\int d^{3}x\;\;\hat{\psi}_{\uparrow}^{\dagger}(\mathbf{x}%
)\hat{\psi}_{\uparrow}^{\dagger}(\mathbf{x})\hat{\psi}_{\uparrow}%
(\mathbf{x})\hat{\psi}_{\uparrow}(\mathbf{x})\right.  \nonumber\\
& +a_{\downarrow}\int d^{3}x\;\;\hat{\psi}_{\downarrow}^{\dagger}%
(\mathbf{x})\hat{\psi}_{\downarrow}^{\dagger}(\mathbf{x})\hat{\psi
}_{\downarrow}(\mathbf{x})\hat{\psi}_{\downarrow}(\mathbf{x})\nonumber\\
& \left.  +2a_{u}\int d^{3}x\;\;\hat{\psi}_{\uparrow}^{\dagger}(\mathbf{x}%
)\hat{\psi}_{\downarrow}^{\dagger}(\mathbf{x})\hat{\psi}_{\downarrow
}(\mathbf{x})\hat{\psi}_{\uparrow}(\mathbf{x})\right)  .\label{b2}%
\end{eqnarray}
By virtue of Pauli exclusion principle, fermion particles occupying two spin states
$\left\vert\uparrow\right\rangle$ and $\left\vert\downarrow\right\rangle$ only interact via short-ranged
interactions if they are in different spin states,
\begin{equation}
\hat{H}_{int} = \left( \frac{4\pi\hbar^{2}a_{F}}{m} \right)
\int d^{3} x \; \;
\hat{\psi}^{\dagger}_{\uparrow}({\bf x}) \hat{\psi}^{\dagger}_{\downarrow}({\bf x})
\hat{\psi}_{\downarrow}({\bf x}) \hat{\psi}_{\uparrow}({\bf x}) \; ,
\label{f2}
\end{equation}
where $a_{F}$ describes the low energy fermion-fermion scattering.  We
obtain these expressions from a scattering picture and derive a pseudo
spin-spin form of the particle-particle effective interaction potentials.

\begin{figure}[pth]
\begin{center}
\includegraphics[width=3in] {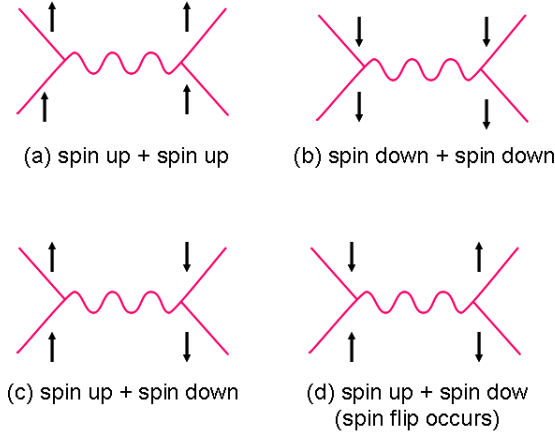}
\end{center}
\caption{Graphic illustration of the four types of binary atom scattering
processes: (a) two spin-up particles collide, (b) two spin down particles
collide, (c) one spin up and one spin-down particle particle scatter while
the spin projection of the interacting particles remains the same, and (d)
a spin-up and a spin-down particle exchange spin states in the collision.}
\label{Fig2collision}
\end{figure}

If the effective pseudo spin projection is conserved, interacting spin $\frac{1}{2}$ atoms, $1$ and $2$,
can undergo four types of scattering events, represented in Fig.~\ref{Fig2collision},
$\left\vert\uparrow\uparrow\right\rangle
\rightarrow\left\vert\uparrow\uparrow \right\rangle$,
$\left\vert\downarrow\downarrow\right\rangle
\rightarrow \left\vert\downarrow\downarrow \right\rangle$,
$\left\vert\uparrow \downarrow\right\rangle
\rightarrow \left\vert\uparrow\downarrow\right\rangle$
(with the same amplitude as
$\left\vert\downarrow\uparrow\right\rangle \rightarrow \left\vert\downarrow\uparrow\right\rangle$
)
and
$\left\vert\downarrow\uparrow\right\rangle\rightarrow \left\vert\uparrow\downarrow\right\rangle$,
(with the same amplitude as
$\left\vert\uparrow\downarrow\right\rangle \rightarrow \left\vert\downarrow\uparrow\right\rangle$
)
where the first arrow in the brackets indicates the spin of atom $1$, the second arrow
shows the spin of atom $2$.
By construction, the effective interaction yields the correct transition matrix elements
in the Born-approximation for the four types of binary atom scattering events:
$(4\pi\hbar^{2}a_{\downarrow}/m)$ for the  mutual scattering of two
`down' particles,
$(4\pi\hbar^{2}a_{\uparrow}/m)$ for the mutual scattering of two
`up' particles,
$(4\pi\hbar^{2}a_{D}/m)$ for direct scattering (unlike spin scattering without spin flip)
and $(4\pi\hbar^{2}a_{x}/m)$ for exchange scattering events (unlike spin scattering
with spin flip).

Denoting the full coordinate of particle $i$, which consists of position
$\mathbf{x}_{i}$ and spin $\mathbf{s}_{i}$ by $\mathbf{r}_{i}$, we can
replace the effective interaction potential Eq.\ (\ref{delta}) by
\begin{equation}
V\left(  \mathbf{r}_{1},\mathbf{r}_{2}\right)  =\frac{4\pi\hbar^{2}}{m}\hat
{a}\delta\left(  \mathbf{x}_{1}-\mathbf{x}_{2}\right)  , \label{delta_12}
\end{equation}
which is a spin-operator.  In the basis of the up and down spins, $\left\{  \left\vert \uparrow\uparrow
\right\rangle ,\left\vert \uparrow\downarrow\right\rangle ,\left\vert
\downarrow\uparrow\right\rangle ,\left\vert \downarrow\downarrow\right\rangle
\right\}  $, the $\hat{a}$-operator reads
\begin{equation}
\hat{a}=\left(
\begin{array}
[c]{llll}
a_{\uparrow} & 0 & 0 & 0\\
0 & a_{D} & a_{x} & 0\\
0 & a_{x} & a_{D} & 0\\
0 & 0 & 0 & a_{\downarrow}
\end{array}
\right)  , \label{amatrix}
\end{equation}
in accordance with the above description.

To connect with spin physics we write the spin dependent scattering length operator,
$\hat{a}$, as the sum of products of single particle
spin-operators.  We write Eq.\ (\ref{amatrix}) as
\begin{equation}
\hat{a}=a_{o}\hat{I}+a_{z,+}\left(  s_{1,z}+s_{2,z}\right)  +a_{z,p}\left[
4\left(  s_{1,z}s_{2,z}\right)  \right]  +a_{x}\left[  2\left(  s_{1,x}%
s_{2,x}+s_{1,y}s_{2,y}\right)  \right],
\end{equation}
where $\hat{I}$ denotes the unit operator. The operators in square
brackets are the spin projection sum operator,
\begin{equation}
s_{1,z}+s_{2,z}=\left(
\begin{array}
[c]{cccc}%
1 & 0 & 0 & 0\\
0 & 0 & 0 & 0\\
0 & 0 & 0 & 0\\
0 & 0 & 0 & -1
\end{array}
\right),
\end{equation}
the spin projection product operator,
\begin{equation}
s_{1,z}s_{2,z}=\left(
\begin{array}
[c]{cccc}
1 & 0 & 0 & 0\\
0 & -1 & 0 & 0\\
0 & 0 & -1 & 0\\
0 & 0 & 0 & 1
\end{array}
\right),
\end{equation}
and the exchange operator,
\begin{equation}
s_{1,x}s_{2,x}+s_{1,y}s_{2,y}=\left(
\begin{array}
[c]{llll}
0 & 0 & 0 & 0\\
0 & 0 & 1 & 0\\
0 & 1 & 0 & 0\\
0 & 0 & 0 & 0
\end{array}
\right).
\end{equation}
By comparing the off-diagonal matrix elements, it is clear that the exchange
operator coefficient must be equal to the exchange scattering length $a_{x}$.
Identifying the diagonal matrix elements we find
$a_{o},a_{z,p},a_{z,+}$;
\begin{eqnarray}
&  a_{o}=\frac{a_{\uparrow}+a_{\downarrow}+2a_{D}}{4},\nonumber\\
&  a_{z,p}=\frac{a_{\uparrow}+a_{\downarrow}-2a_{D}}{4},\nonumber\\
&  a_{z,+}=\frac{a_{\uparrow}-a_{\downarrow}}{2} \; .
\end{eqnarray}
Hence, the expression
\begin{eqnarray}
V(\vec{r}_{1},\vec{r}_{2}) &=&\left( \frac{4\pi \hbar ^{2}}{m}\right) \delta
(\mathbf{x}_{1}-\mathbf{x}_{2})\left\{ \left[ \frac{a_{\uparrow
}+a_{\downarrow }+2a_{D}}{4}\right] +\left[ \frac{a_{\uparrow
}-a_{\downarrow }}{2}\right] \left( s_{1,z}+s_{2,z}\right) \right.
\nonumber\\
&&\left. +\left[ a_{\uparrow }+a_{\downarrow }-2a_{D}\right] \left(
s_{1,z}s_{2,z}\right) +2a_{x}\left( s_{1,x}s_{2,x}+s_{1,y}s_{2,y}\right)
\right\}.
\label{effspin}
\end{eqnarray}
translates the effective interaction potential operator Eq.\ (\ref{delta_12})
into pseudo-spin language.

\subsection{A magnetic-like interaction form of the inter-particle interactions}

In the interest of exploiting the magnetism analogy, we note that the effective
spin-dependent interaction Eq.\ (\ref{effspin}) can be rephrased as a magnetic
interaction.  Specifically, the spin-dependent part stems from the interaction
of the effective spin of one particle with an effective short-ranged magnetic
field carried by the other particle.  That description allows one to interpret the
interaction of one particle in a many-body system to be caused by the interaction with
an external field and with an internal magnetic field generated by the
moments of the other particles.

We write the effective interaction of atoms $1$ and $2$, as the sum
of a spin independent contact interaction, $V_{o}$ and and a spin
interaction, $V_{s}$, $V\left(\vec{r}_{1},\vec{r}_{2}\right)=V_{o}
\left(\vec{x}_{1}-\vec{x}_{2}\right)+V_{s}\left(\vec{r}_{1},\vec{r}_{2}\right)$ where
\begin{equation}
V_{o}\left(  \mathbf{x}_{1}-\mathbf{x}_{2}\right)  =\frac{\pi\hbar^{2}}
{m}\left(  a_{\uparrow}+a_{\downarrow}+2a_{D}\right)  \delta\left(
\mathbf{x}_{1}-\mathbf{x}_{2}\right).
\label{spinin2}
\end{equation}
The spin-dependent part we cast into the form of the energy of a spin in a magnetic
field with unit effective Bohr-magneton (as we chose the spin to be
dimensionless, the effective magnetic field than takes the units of energy)
\begin{equation}
V_{s}\left(  \mathbf{r}_{1},\mathbf{r}_{2}\right)  =\mathbf{s}_{1}
\cdot\mathbf{h}_{2}\left(  \mathbf{x}_{1}\right)  +\mathbf{s}_{2}
\cdot\mathbf{h}_{1}\left(  \mathbf{x}_{2}\right)  , \label{spin2}
\end{equation}
where we assume the effective $\mathbf{h}$--magnetic field to be
the short-range part of an effective moment $\mathbf{m}$,
\begin{equation}
\mathbf{h}_{i}\left(  \mathbf{x}\right)  =\mathbf{m}_{i}\delta\left(
\mathbf{x}-\mathbf{x}_{i}\right)  .\; \label{h2}
\end{equation}
The effective $\mathbf{m}$-moment is an operator,
\begin{equation}
\mathbf{m}_{i}=m_{s,o}\mathbf{z}+m_{s,\parallel}\mathbf{z}\;\left(
\;\mathbf{s}_{i}\;\cdot\mathbf{z}\;\right)  +m_{s,\perp}\;\mathbf{z}
\times\left(  \;\mathbf{s}_{i}\times\mathbf{z}\;\right)  , \label{mom2}
\end{equation}
with moment parameters
\begin{eqnarray}
m_{s,o}  &  =\frac{2\pi\hbar^{2}}{m}\left(  a_{\uparrow}-a_{\downarrow
}\right)  ,\nonumber\\
m_{s,\parallel}  &  =\frac{2\pi\hbar^{2}}{m}\left(  a_{\uparrow}%
+a_{\downarrow}-2a_{D}\right)  ,\nonumber\\
m_{s,\perp}  &  =\frac{4\pi\hbar^{2}}{m}a_{x}, \label{gyro2}
\end{eqnarray}
that depend on the scattering lengths.

\subsection{The effect of quantum statistics on short-range spin-spin interactions}

As the pseudo spin operators are not the generators of rotations, the spin-spin
interaction can be and generally is anisotropic.  The $m_{o}$-term describes
the interaction of $\vec{s}$ with an effective
short-range magnetic moment that points in the $\vec{z}$-direction (the direction of the physical
magnetic field, the actual field that splits the Zeeman-levels) and is
independent of the pseudo-spin of the particle that carries the moment.
Another source of anisotropy is the difference in the $m_{\parallel}$ and
$m_{\perp}$-moment parameters.  The short-range nature of the interactions and
the quantum statistics of the interacting particles, give a moment that can
be chosen to align itself with the magnetic field direction or to be
perpendicular to it.  To see that, consider the interaction
\begin{equation}
\hat{H}_{int}=\frac{1}{2}\left(  \frac{4\pi\hbar^{2}}{m}\right)  \int
d^{3}x\;\;\left\langle \hat{\psi}^{\dagger}\left(  \vec{x}\right)  \hat{\psi
}^{\dagger}\left(  \vec{x}\right)  \left\vert \hat{a}\right\vert \hat{\psi
}\left(  \vec{x}\right)  \hat{\psi}\left(  \vec{x}\right)  \right\rangle,
\end{equation}%
where $\hat{a}$ denotes the spin-dependent scattering length operator of Eq.\ (%
\ref{amatrix}), where
\begin{equation}
\left. |\hat{\psi}\left( \vec{x}\right) \hat{\psi}\left( \vec{x}\right)
\right\rangle =\left(
\begin{array}{l}
\hat{\psi}_{\uparrow }\left( \vec{x}\right) \hat{\psi}_{\uparrow }\left(
\vec{x}\right)  \\
\hat{\psi}_{\uparrow }\left( \vec{x}\right) \hat{\psi}_{\downarrow }\left(
\vec{x}\right)  \\
\hat{\psi}_{\downarrow }\left( \vec{x}\right) \hat{\psi}_{\uparrow }\left(
\vec{x}\right)  \\
\hat{\psi}_{\downarrow }\left( \vec{x}\right) \hat{\psi}_{\downarrow }\left(
\vec{x}\right)
\end{array}%
\right),
\label{2spin}
\end{equation}%
indicates the two-particle spinor.
The short-range nature of the interaction causes the field operators to be evaluated at
the same position, $\vec{x}$.  As pairs of annihilation and creation operators of the same argument,
the two-particle spinor components obey commutator (anti-commutator) relations if the interacting
particles are bosonic (fermionic), $\hat{\psi}_{i}\left(\vec{x}\right) \hat{\psi}_{j}\left(\vec{x}\right)
= \pm \hat{\psi}_{j}\left(\vec{x}\right) \hat{\psi}_{i}\left(\vec{x}\right)$.  By writing the components of the
two particle spinor Eq.\ (\ref{2spin}) as half the sum with itself, then replacing the second term by
$\pm$ its reverse order, we obtain, for bosons
\begin{equation}
\left\vert \hat{\psi}\left(  \vec{x}\right)  \hat{\psi}\left(  \vec{x}\right)
\right\rangle =\left(
\begin{array}
[c]{c}%
\hat{\psi}_{\uparrow}\left(  \vec{x}\right)  \hat{\psi}_{\uparrow}\left(
\vec{x}\right)  \\
\frac{1}{\sqrt{2}}\left(  \frac{\hat{\psi}_{\uparrow}\left(  \vec{x}\right)
\hat{\psi}_{\downarrow}\left(  \vec{x}\right)  +\hat{\psi}_{\downarrow}\left(
\vec{x}\right)  \hat{\psi}_{\uparrow}\left(  \vec{x}\right)  }{\sqrt{2}%
}\right)  \\
\frac{1}{\sqrt{2}}\left(  \frac{\hat{\psi}_{\downarrow}\left(  \vec{x}\right)
\hat{\psi}_{\uparrow}\left(  \vec{x}\right)  +\hat{\psi}_{\uparrow}\left(
\vec{x}\right)  \hat{\psi}_{\downarrow}\left(  \vec{x}\right)  }{\sqrt{2}%
}\right)  \\
\hat{\psi}_{\downarrow}\left(  \vec{x}\right)  \hat{\psi}_{\downarrow}\left(
\vec{x}\right)
\end{array}
\right)  ,\label{bspin}%
\end{equation}
for bosons and for fermions
\begin{equation}
\left\vert \hat{\psi}\left(  \vec{x}\right)  \hat{\psi}\left(  \vec{x}\right)
\right\rangle =\left(
\begin{array}
[c]{c}%
0\\
\frac{1}{\sqrt{2}}\left(  \frac{\hat{\psi}_{\uparrow}\left(  \vec{x}\right)
\hat{\psi}_{\downarrow}\left(  \vec{x}\right)  -\hat{\psi}_{\downarrow}\left(
\vec{x}\right)  \hat{\psi}_{\uparrow}\left(  \vec{x}\right)  }{\sqrt{2}%
}\right)  \\
\frac{1}{\sqrt{2}}\left(  \frac{\hat{\psi}_{\downarrow}\left(  \vec{x}\right)
\hat{\psi}_{\uparrow}\left(  \vec{x}\right)  -\hat{\psi}_{\uparrow}\left(
\vec{x}\right)  \hat{\psi}_{\downarrow}\left(  \vec{x}\right)  }{\sqrt{2}%
}\right)  \\
0
\end{array}
\right)  .\label{fspin}%
\end{equation}
The resulting non-vanishing column matrix elements are components of the
pseudo-spin triplet manifold in the boson-case and the pseudospin singlet state
in the fermion case.  We find that the two-particle spinor is projected onto the triplet
subspace if the fields are bosonic and onto the singlet subspace if the fields are fermionic,
\begin{eqnarray}
\left\vert \hat{\psi}\left( \vec{x}\right) \hat{\psi}\left( \vec{x}\right)
\right\rangle _{bosons} &=&\hat{\Pi}_{T}\left\vert \hat{\psi}\left( \vec{x}%
\right) \hat{\psi}\left( \vec{x}\right) \right\rangle ,
\nonumber\\
\left\vert \hat{\psi}\left( \vec{x}\right) \hat{\psi}\left( \vec{x}\right)
\right\rangle _{fermions} &=&\hat{\Pi}_{S}\left\vert \hat{\psi}\left( \vec{x}%
\right) \hat{\psi}\left( \vec{x}\right) \right\rangle ,
\end{eqnarray}
where the $\hat{\Pi}_{T}$ ($\hat{\Pi}_{S}$) project onto the two-particle
pseudospin triplet (singlet) subspace.  We can check that statement by direct inspection.
For instance, the second component of the column matrices Eq.\ (\ref{bspin}) and Eq.\ (\ref{fspin})
represents the two-particle annihilation field of the $\vert\uparrow \downarrow\rangle$--state,
which can be written as
$|\uparrow \downarrow \rangle =\frac{1}{\sqrt{2}}\left( \frac{%
|\uparrow \downarrow \rangle +|\downarrow \uparrow \rangle }{\sqrt{2}}+\frac{%
|\uparrow \downarrow \rangle -|\downarrow \uparrow \rangle }{\sqrt{2}}%
\right) $,
so that
$\hat{\Pi}_{T}|\uparrow \downarrow \rangle =\frac{1}{%
\sqrt{2}}\frac{|\uparrow \downarrow \rangle +|\downarrow \uparrow \rangle }{%
\sqrt{2}}$ and $\hat{\Pi}_{S}|\uparrow \downarrow \rangle =\frac{1}{\sqrt{2}}%
\frac{|\uparrow \downarrow \rangle -|\downarrow \uparrow \rangle }{\sqrt{2}}$%
, corresponding to the spin states of the second component of the right-hand
sides of Eqs.\ (\ref{bspin}) and (\ref{fspin}).

Writing the boson matrix product of the interaction Hamiltonian out, we find that we can also
write the effective interaction Hamiltonian as a bracket of the triplet two-particle spinor,
\begin{equation}
\left\vert \hat{\psi}\left( \vec{x}\right) \hat{\psi}\left( \vec{x}\right)
\right\rangle _{T}=\left(
\begin{array}{c}
\hat{\psi}_{\uparrow }\left( \vec{x}\right) \hat{\psi}_{\uparrow }\left(
\vec{x}\right)  \\
\frac{\hat{\psi}_{\uparrow }\left( \vec{x}\right) \hat{\psi}_{\downarrow
}\left( \vec{x}\right) +\hat{\psi}_{\downarrow }\left( \vec{x}\right) \hat{%
\psi}_{\uparrow }\left( \vec{x}\right) }{\sqrt{2}} \\
\hat{\psi}_{\downarrow }\left( \vec{x}\right) \hat{\psi}_{\downarrow }\left(
\vec{x}\right)
\end{array}%
\right),
\label{tspin}
\end{equation}%
in terms of which
\begin{equation}
\hat{H}_{int}=\frac{1}{2}\left( \frac{4\pi \hbar ^{2}}{m}\right) \int
d^{3}x\;\;_{T}\left\langle \hat{\psi}^{\dagger }\left( \vec{x}\right) \hat{%
\psi}^{\dagger }\left( \vec{x}\right) |\hat{a}_{T}|\hat{\psi}\left( \vec{x}%
\right) \hat{\psi}\left( \vec{x}\right) \right\rangle _{T},
\label{bint}
\end{equation}%
where the triplet scattering length operator $\hat{a}_{T}$ is now
represented by a diagonal matrix,
\begin{equation}
\hat{a}=\left(
\begin{array}{lll}
a_{\uparrow } & \;\;\;\;\;0 & 0 \\
0 & a_{D}+a_{x} & 0 \\
0 & \;\;\;\;\;0 & a_{\downarrow }%
\end{array}%
\right).
\label{atmatrix}
\end{equation}%
One more application of the commutator relations casts the interaction Hamiltonian
Eq.\ (\ref{bint}) in the form of Eq.\ (\ref{b2}) provided we identify the unlike boson scattering length
$a_{u}$ with
$a_{u}=a_{D}+a_{x}$.  Likewise, the fermion interaction
\begin{equation}
\hat{H}_{int}=\frac{1}{2}\left( \frac{4\pi \hbar ^{2}}{m}\right) \int
d^{3}x\;\;_{S}\left\langle \hat{\psi}^{\dagger }\left( \vec{x}\right) \hat{%
\psi}^{\dagger }\left( \vec{x}\right) |\hat{a}_{S}|\hat{\psi}\left( \vec{x}%
\right) \hat{\psi}\left( \vec{x}\right) \right\rangle _{S},
\label{fint}
\end{equation}
with two-particle singlet spin component
\begin{equation}
\left\vert \hat{\psi}\left( \vec{x}\right) \hat{\psi}\left( \vec{x}\right)
\right\rangle _{S}=\frac{\hat{\psi}_{\uparrow }\left( \vec{x}\right) \hat{%
\psi}_{\downarrow }\left( \vec{x}\right) -\hat{\psi}_{\downarrow }\left(
\vec{x}\right) \hat{\psi}_{\uparrow }\left( \vec{x}\right) }{\sqrt{2}}\;,
\end{equation}
and singlet scattering length,
\begin{equation}
\hat{a}_{S}=a_{D}-a_{x} \; ,
\end{equation}
reduces to the form of Eq.\ (\ref{f2}) if we identify $a_{F}$ with $a_{F}=a_{D} - a_{x}$.

As an interesting consequence of the triplet projection caused by boson statistics,
the spin-spin interaction can be written in two equivalent forms. As $\vec{s}_{1}
\cdot\vec{s}_{2} =1/4$ in a triplet state, we can either replace
$s_{1,z}s_{2,z}=1/4-\left(s_{1,x}s_{2,x}+s_{1,y}s_{2,y}\right)$ or,
alternatively, $s_{1,x}s_{2,x}+s_{1,y}s_{2,y}$ by  $s_{1,x}s_{2,x}+s_{1,y}s_{2,y}=1/4-s_{1,z}s_{2,z}$.
As a consequence of the second replacement, the effective spin-spin interaction
takes the form of a short-range Ising-like interaction, the second replacement gives
a short-range XY spin-spin interaction.  The resulting boson spin-spin interactions
\begin{eqnarray}
V_{I}\left( \vec{r}_{1},\vec{r}_{2}\right)  &=&\left( \frac{4\pi \hbar ^{2}}{%
m}\right) \delta \left( \vec{x}_{1}-\vec{x}_{2}\right) \left\{ \left[ \frac{%
a_{\uparrow }+a_{\downarrow }+2a_{u}}{4}\right] \right.
\nonumber\\
&&\left. +\left[ \frac{a_{\uparrow }-a_{\downarrow }}{2}\right] \left(
s_{1,z}+s_{2,z}\right) +\left[ a_{\uparrow }+a_{\downarrow }-2a_{u}\right]
\left( s_{1,z}s_{2,z}\right) \right\},
\label{I}
\end{eqnarray}%
and
\begin{eqnarray}
V_{XY}\left(  \vec{r}_{1},\vec{r}_{2}\right)    & =\left(  \frac{4\pi\hbar
^{2}}{m}\right)  \delta\left(  \vec{x}_{1}-\vec{x}_{2}\right)  \left\{
\left[  \frac{a_{\uparrow}+a_{\downarrow}}{2}\right]  +\left[  \frac
{a_{\uparrow}-a_{\downarrow}}{2}\right]  \left(  s_{1,z}+s_{2,z}\right)
\right.  \nonumber\\
& \;\left.  -\left[  a_{\uparrow}+a_{\downarrow}-2a_{u}\right]  \left(
s_{1,x}s_{2,x}+s_{1,y}s_{2,y}\right)  \right\}
\label{XY}%
\end{eqnarray}
are equivalent to each other and to the more conventional expression of Eq.\ (\ref{b2}).
One advantage of the $XY$-form, Eq.\ (\ref{XY}), $V_{XY}\left(\vec{r}_{1},\vec{r}_{2}\right)$,
is that the unlike boson scattering length $a_{u}$ only occurs in the spin-spin term.  As
a consequence a mixed spin channel Feshbach resonance will vary only the
$XY$spin-spin coupling.  In general, a Feshbach resonance occurs when the
incident particle channel becomes degenerate with the quasi-bound state of
another collision channel.  Hence a particular resonance will either vary $a_{\uparrow}$,
or $a_{\downarrow}$ or $a_{u}$.  A mixed spin channel resonance of magnetic
field width $\Delta B$ around magnetic field strength $B_{res}$ varies the unlike
scattering length $a_{u}$ as
\begin{equation}
a_{u,res}\left(B\right)=a_{u}\left[ 1 - \frac{\Delta B}{B-B_{res}} \right] \; ,
\end{equation}
where $a_{u}$ is the background scattering length that varies slowly
with magnetic field (on the magnetic field scale of $B_{hf}$.  The Feshbach
variation thereby adds an effective interaction potential
\begin{equation}
\Delta V_{XY}\left(\vec{r}_{1},\vec{r}_{2}\right) = \left(\frac{\Delta B}{B_{res}-B}\right)
\left( \frac{4\pi\hbar^{2} a_{u}}{m} \right) \delta\left(\vec{x}_{1}-\vec{x}_{2}\right)
 \left(s_{1,x}s_{2,x}+s_{1,y}s_{2,y}\right)
\end{equation}
to the above $XY$ form of the spin-dependent particle-particle interaction Eq.\ (\ref{XY}).

In terms of the magnetic form of the inter-particle interactions,
\begin{equation}
V_{I(X,Y)}=V_{o,I(X,Y)}\left(\vec{x}_{1}-\vec{x}_{2}\right)
+V_{s,I(X,Y)}\left(\vec{r}_{1},\vec{r}_{2}\right),
\end{equation}
where $V_{o,I(XY)}=[4\pi\hbar^{2}/m] \delta\left(\vec{x}_{1}-\vec{x}_{2}\right)
a_{o,I(X,Y)}$ in which
\begin{eqnarray}
a_{o,I}&=&\frac{a_{\uparrow}+a_{\downarrow}+2a_{u}}{4},
\nonumber \\
a_{o,XY}&=&\; \; \; \frac{a_{\uparrow}+a_{\downarrow}}{2},
\end{eqnarray}
denote the Ising and XY expressions of the spin independent scattering length.
The spin interactions take the usual form $V_{s,I(XY)}
\left(\vec{r}_{1},\vec{r}_{2}\right)=\mathbf{s}_{1}\cdot\mathbf{h}_{2,I(XY)}\left(\vec{x}_{1}\right)
+ \mathbf{s}_{2}\cdot\mathbf{h}_{1,I(XY)}\left(\vec{x}_{2}\right)$ with $\mathbf{h}_{i,I(XY)}
\left(\vec{x}\right)=\delta\left(\vec{x}-\vec{x}_{i}\right) \mathbf{m}_{i,I(XY)}$ and
\begin{eqnarray}
\mathbf{m}_{i,I} &=& m_{o} \mathbf{z} + \left( m_{\parallel}-m_{\perp} \right)
\mathbf{z} \; \left( \mathbf{s}_{i} \cdot \mathbf{z} \right),
\nonumber \\
\mathbf{m}_{i,XY} &=& m_{o} \mathbf{z} + \left( m_{\perp}-m_{\parallel} \right)
\mathbf{z} \times \left( \mathbf{s}_{i} \times \mathbf{z} \right),
\end{eqnarray}
with moment parameters ($m_{o},m_{\parallel},m_{\perp}$) defined in Eq.\ (\ref{gyro2}).

\subsection{Spin dependence of alkali atom interactions and the degenerate internal state approximation}

The spin-dependence of the alkali-atom interactions stems from the exchange of
valence electrons.
As two alkali-nuclei approach each other to nanometer and sub-nanometer distance,
their valence electrons, now encircling the two closely-spaced nuclei, become
strongly correlated. If these electrons are indistinguishable, i.e., if their
spins align in a triplet sate, the likelihood of finding them simultaneously
in each others vicinity is reduced by virtue of the Pauli principle. Pauli
exclusion then reduces the Coulomb energy shift in the inter-atomic potential.
In contrast, spin singlet electrons can approach each other more closely,
shifting the inter-atomic potential upward. Hence, the triplet potential $V_{T}$
is generally deeper than the singlet potential $V_{S}$. The overall interaction of
atoms can be expressed by an inter-atomic potential operator
\begin{equation}
V\left(  \mathbf{r}_{1},\mathbf{r}_{2}\right)  =V_{T}\left(  \left\vert
\mathbf{x}_{1}-\mathbf{x}_{2}\right\vert \right)  \hat{\Pi}_{e,T}+V_{S}\left(
\left\vert \mathbf{x}_{1}-\mathbf{x}_{2}\right\vert \right)  \hat{\Pi}_{e,S},
\label{micrint}%
\end{equation}
where the $\mathbf{r}$ represent both the spatial coordinates $\mathbf{x}$,
and spin, and $\hat{\Pi}_{e,T}$ , $\hat{\Pi}_{e,S}$ denote the projection
operators for the electron triplet $S=1$ and singlet $S=0$ states. When acting
upon a triplet state, the square of the total electron spin operator,
$\mathbf{S}_{e}^{2}=\frac{3}{2}+2\mathbf{s}_{e,1}\cdot\mathbf{s}_{e,2}$ yields
an eigenvalue of $2$ (i.e. $S\left(  S+1\right)  $ with $S=1$). Acting upon a
singlet state, the same operator gives zero, so that the triplet projection
operator takes the form
\begin{equation}
\hat{\Pi}_{e,T}=\frac{\mathbf{S}_{e}^{2}}{2}=\frac{3}{4}+\mathbf{s}_{e,1}
\cdot\mathbf{s}_{e,2}. \label{proj1}
\end{equation}
Since $\hat{\Pi}_{e,T}+\hat{\Pi}_{e,S}=1$, the singlet projection operator is
equal to
\begin{equation}
\hat{\Pi}_{e,S}=1-\hat{\Pi}_{e,T}=\frac{1}{4}-\mathbf{s}_{e,1}
\cdot\mathbf{s}_{e,2} \; . \label{proj2}
\end{equation}
With Eqs.\ (\ref{proj1}) and (\ref{proj2}), the interatomic interaction
potential operator Eq.\ (\ref{micrint}) reads
\begin{eqnarray}
&  V\left(  \mathbf{r}_{1},\mathbf{r}_{2}\right)  =\frac{1}{4}\left[
3V_{T}\left(  \left\vert \mathbf{x}_{1}-\mathbf{x}_{2}\right\vert \right)
+V_{S}\left(  \left\vert \mathbf{x}_{1}-\mathbf{x}_{2}\right\vert \right)
\right] \nonumber\\
&  \;\;\;\;\;+\left[  V_{T}\left(  \left\vert \mathbf{x}_{1}-\mathbf{x}
_{2}\right\vert \right)  -V_{S}\left(  \left\vert \mathbf{x}_{1}
-\mathbf{x}_{2}\right\vert \right)  \right]  \mathbf{s}_{e,1}\cdot
\mathbf{s}_{e,2} \;.
\label{sintrip}
\end{eqnarray}
The depth of the $V_{S}$ and $V_{T}$ potentials ($\sim$ electron Volt - $10^{4} K$)
greatly exceeds the energy of the Zeeman spin interactions $\sim a_{hf}$, tens of $mK$).
Inside the potential well, $r < r_{1}$, the hyperfine interaction can be neglected
whereas in the outer region, $r > r_{1}$, the hyperfine interaction determines the
spin state of the collision channel.  The exchange interaction, $\left(
V_{T}-V_{S} \right) \vec{s}_{e,1}\cdot \vec{s}_{e,2}$ also falls off rapidly
(exponentially) in the outer region.  One can then calculate the scattering wavefunction
while omitting the Zeeman terms in $r < r_{1}$ and treating the exchange interaction
in the region $r>r_{1}$ as a perturbation term.  In lowest order perturbation the
T-matrix should then have a contribution proportional to $\vec{s}_{e,1} \cdot \vec{s}_{e,2}$
evaluated for the initial and outgoing channels.  The spin-independent and
the $\vec{s}_{e,1}\cdot\vec{s}_{e,2}$-parts of the T-matrix should reproduce the
correct triplet and singlet scattering lengths $a_{S}$ and $a_{T}$ in the
limit of vanishing magnetic field and hyperfine energy.  In this approximation,
the low energy $i,j \rightarrow k,l$ transition matrix element then takes the
form,
\begin{equation}
T_{i,j;k,l} \approx \left(\frac{4\pi\hbar^{2}}{m}\right)
\langle i, j \vert \overline{a} + a_{-} \vec{s}_{e,1} \cdot \vec{s}_{e,2}
\vert k,s \rangle \; ,
\end{equation}
where $\overline{a}$ denotes the scattering length averaged over the
singlet and triplet states,
\begin{equation}
\overline{a} = \left( \frac{ 3 a_{T} + a_{S} }{4} \right) \; ,
\end{equation}
and where $a_{-}$ represents the difference scattering length,
\begin{equation}
a_{-} = a_{T} - a_{S} \; \; \; .
\end{equation}
This approximation is called the Degenerate Internal State (DIS) approximation.
While it was primarily designed for calculating the two-body loss rate of
atoms occupying specific hyperfine states in an external magnetic field.  In that
case, if the magnetic field is comparable to the hyperfine field, the above formula
does not work very well as the wavefunction of the outgoing channel is not very
well approximated by the zero energy wave function.  We are not considering
lossy channels, which will not be there if the hyperfine states are chosen
carefully as described in the previous section. We are only considering
the cases $i=k, j=l$ or $i=l, j=k$ with $i,j = \uparrow,\downarrow$, for which the
incident and final channel wavefunctions have the same low energy value.
Even in that case, the approximation is not always satisfied, particularly for
atoms that have a naturally large scattering length at zero magnetic field, such
as $^{7}Li$ \cite{Stoof1} and $^{123}$Cs \cite{Stoof2} and for higher magnetic
field values.  We expect that the treatment, may, however, yield a reasonable
approximation for magnetic field values near such low magnetic field resonances as
the $9 G$ resonance observed in $^{87}Rb $ \cite{Sengstock}.

In the DIS approximation, we can calculate the above defined scattering lengths
explicitly
\begin{eqnarray}
a_{\uparrow} &  =\bar{a}+a_{-}\left\langle \uparrow\uparrow\left\vert
\mathbf{s}_{e,1}\cdot\mathbf{s}_{e,2}\right\vert \uparrow\uparrow\right\rangle
,\nonumber\\
a_{\downarrow} &  =\bar{a}+a_{-}\left\langle \downarrow\downarrow\left\vert
\mathbf{s}_{e,1}\cdot\mathbf{s}_{e,2}\right\vert \downarrow\downarrow
\right\rangle ,\nonumber\\
a_{D} &  =a_{-}\left\langle \uparrow\downarrow\left\vert \mathbf{s}_{e,1}%
\cdot\mathbf{s}_{e,2}\right\vert \uparrow\downarrow\right\rangle ,\nonumber\\
a_{x} &  =a_{-}\left\langle \uparrow\downarrow\left\vert \mathbf{s}_{e,1}%
\cdot\mathbf{s}_{e,2}\right\vert \downarrow\uparrow\right\rangle .\label{ax}%
\end{eqnarray}
in terms of the single electron spin matrix elements $\mathbf{s}_{e,\uparrow
}=\left\langle \uparrow\left\vert \mathbf{s}_{e}\right\vert \uparrow
\right\rangle $, $\mathbf{s}_{e,\downarrow}=\left\langle \downarrow\left\vert
\mathbf{s}_{e}\right\vert \downarrow\right\rangle $,
\begin{eqnarray}
\left\langle \uparrow\uparrow\left\vert \mathbf{s}_{e,1}\cdot\mathbf{s}%
_{e,2}\right\vert \uparrow\uparrow\right\rangle  &  =\mathbf{s}_{e,\uparrow
}\cdot\mathbf{s}_{e,\uparrow},\\
\left\langle \downarrow\downarrow\left\vert \mathbf{s}_{e,1}\cdot
\mathbf{s}_{e,2}\right\vert \downarrow\downarrow\right\rangle  &
=\mathbf{s}_{e,\downarrow}\cdot\mathbf{s}_{e,\downarrow},\\
\left\langle \uparrow\downarrow\left\vert \mathbf{s}_{e,1}\cdot\mathbf{s}%
_{e,2}\right\vert \uparrow\downarrow\right\rangle  &  =\mathbf{s}_{e,\uparrow
}\cdot\mathbf{s}_{e,\downarrow}.
\end{eqnarray}
The exchange spin matrix element involves spin-flip matrix elements of the
type $\left\langle \uparrow\left\vert \mathbf{s}_{e}\right\vert \downarrow
\right\rangle =\mathbf{s}_{e,\uparrow\downarrow}$, so that
\begin{equation}
\left\langle \uparrow\downarrow\left\vert \mathbf{s}_{e,1}\cdot\mathbf{s}%
_{e,2}\right\vert \downarrow\uparrow\right\rangle =\mathbf{s}_{e,\uparrow
\downarrow}\cdot\mathbf{s}_{e,\downarrow\uparrow}\;.
\end{equation}
The spin-spin interaction parameters,
\begin{eqnarray}
a_{o,I} &  =\overline{a}+\frac{a_{-}}{4}\left(  \mathbf{s}_{e,\uparrow
}+\mathbf{s}_{e,\downarrow}\right)  \cdot\left(  \mathbf{s}_{e,\uparrow
}+\mathbf{s}_{e,\downarrow}\right)  ,\nonumber\\
m_{s,o} &  =\frac{2\pi\hbar^{2}}{m}a_{-}\left(  \mathbf{s}_{e,\uparrow
}+\mathbf{s}_{e,\downarrow}\right)  \cdot\left(  \mathbf{s}_{e,\uparrow
}-\mathbf{s}_{e,\downarrow}\right)  ,\nonumber\\
m_{s,\parallel} &  =\frac{2\pi\hbar^{2}}{m}a_{-}\left(  \mathbf{s}%
_{e,\uparrow}-\mathbf{s}_{e,\downarrow}\right)  \cdot\left(  \mathbf{s}%
_{e,\uparrow}-\mathbf{s}_{e,\downarrow}\right)  ,\nonumber\\
m_{s,\perp} &  =\frac{2\pi\hbar^{2}}{m}a_{-}2\mathbf{s}_{e,\uparrow\downarrow
}\cdot\mathbf{s}_{e,\downarrow\uparrow},
\end{eqnarray}
then depend on the single electron spin-flip and spin matrix elements.

As the electron spin expectation values depend on the external magnetic field,
the interaction parameters do as well. To express the dependence explicitly,
we cast the expressions in parametric form, choosing the
electron spin inclination angles $\theta_{\uparrow}$, $\theta_{\downarrow}$ of the
`up' and `down' hyperfine states of Eq.\ (\ref{inclin}) as variable. For notational convenience
we introduce external magnetic field-dependent spin factors $\mathcal{S}_{m,n}(b)$ with
$m$ and $n$ equal to $+1$ or $-1$,
\begin{eqnarray}
\mathcal{S}_{m,n}\left(  b\right)    & =\left(  \mathbf{s}_{e,\uparrow}%
+m\vec{s}_{e,\downarrow}\right)  \cdot\left(  \mathbf{s}_{e,\uparrow
}+n\mathbf{s}_{e,\downarrow}\right),\nonumber\\
& =\left[  \frac{\cos\left(  \theta_{\uparrow}\right)  +m\cos\left(
\theta_{\downarrow}\right)  }{2}\right]  \left[  \frac{\cos\left(
\theta_{\uparrow}\right)  +n\cos\left(  \theta_{\downarrow}\right)  }%
{2}\right].
\end{eqnarray}
In addition, we introduce the exchange spin factor, $\mathcal{S}_{x}\left(
b\right)  $, with $\mathcal{S}_{x}\left(  b\right)  =2\mathbf{s}
_{e,\uparrow\downarrow}\cdot\mathbf{s}_{e,\downarrow\uparrow}$. To determine
its value we make assumptions about how the $\left\vert \uparrow
\right\rangle $ and $\left\vert \downarrow\right\rangle $ are chosen: We assume that their
respective $m_{f}$-values differ by one unit (if not, $\mathbf{s}
_{e,\uparrow\downarrow}=\mathbf{s}_{e,\downarrow\uparrow}=0$) and we choose
the $\left\vert \uparrow\right\rangle $ to have the highest $m_{f}$.
We also assume that both states are chosen among the states
with Zeeman energy-curves that slope down at high magnetic fields
(either the $f^{-}$-states or the stretched electron spin-down state, $\left\vert f=f^{+}
,m_{f}=-f^{+}\right\rangle $).  In that case the $\left\vert \uparrow
\right\rangle $-state has the lower Zeeman energy. With this
convention, we find
\begin{equation}
\mathcal{S}_{x}\left(  b\right)  =\left[  \frac{1+\cos\left(  \theta
_{\uparrow}\right)  }{2}\right]  \left[  \frac{1-\cos\left(  \theta
_{\downarrow}\right)  }{2}\right]  ,
\end{equation}
where the projection of the inclination angle varies with the external
magnetic field as in Eq.~(\ref{inclin}). In terms of the $\mathcal{S}$--spin
factors, the effective pseudospin interaction parameters take on simple forms
\begin{eqnarray}
a_{o,I} &=& \bar{a}+\frac{a_{-}}{4} \;  \mathcal{S}_{+,+}\left(  b\right)  \; ,
\nonumber \\
m_{s,o}  &=& \frac{2\pi\hbar^{2}}{m}a_{-}\;\mathcal{S}_{+,-}\left(  b\right) \;
,\nonumber\\
m_{s,\parallel}  &=& \frac{2\pi\hbar^{2}}{m}a_{-}\;\mathcal{S}_{-,-}\left(
b\right) \;  ,\nonumber\\
m_{s,\perp}  & =& \frac{2\pi\hbar^{2}}{m}a_{-}\;\mathcal{S}_{x}\left(
b\right)  \; \; .
\end{eqnarray}
Note that all three moment parameters are proportional to the
difference scattering length $a_{-}$.

\begin{figure}[ptb]
\begin{center}
\includegraphics[width=2.8in] {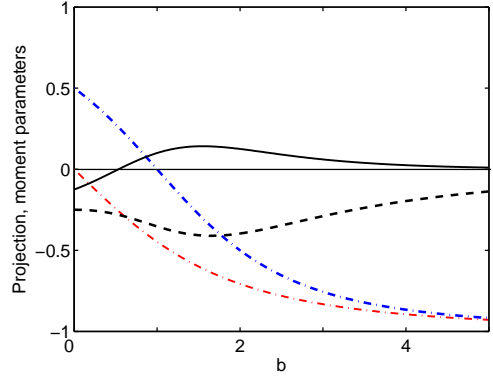}
\end{center}
\caption{The graph shows the magnetic field dependence
($b=B/B_{hf}$) of the moment parameters as calculated in the DIS
approximation for $i=3/2$, $\left\vert \uparrow\right\rangle=\vert f=1,m_{f}=0\rangle$
and $\vert f=1,m_{f}=-1\rangle$.  The full line shows
$m_{0}/\left(  \pi\hbar^{2}/m\right) a_{-}  $,
whereas the dashed line shows
$2\left(  m_{\parallel}-m_{\perp
}\right)  /\left(  \pi\hbar^{2}/m\right)  a_{-}  $,
where $a_{-}$ represents the difference of the zero magnetic field triplet
($a_{T}$) and singlet ($a_{S}$) scattering lengths, $a_{-}=a_{T}-a_{S}$.
The dashed lines show the cosine of the electron spin projection
angle of the `up' and `down' hyperfine states.  The higher lying dashed
line plots $\cos\left(\theta_{\downarrow}\right)$, the lower dashed line
plots $\cos\left(\theta_{\uparrow}\right)$.  It is near the $b=1$ at
which the down state electron spin changes its direction from parallel to
antiparellel to the external magnetic field that $m_{o}$ changes sign and
$\vert m_{\parallel}-m_{\perp} \vert $ is maximized.}
\label{Fig3moment}
\end{figure}

Figure\ \ref{Fig3moment} plots the magnetic field dependence
($b=B/B_{hf}$ as defined above Eq.\ (\ref{zeeman})) of the
relevant moment-parameters as calculated in the
DIS approximation for the special case,
$i=3/2$, $\left\vert \uparrow\right\rangle=\vert f=1,m_{ff}=0\rangle$
and $\vert f=1,m_{f}=-1\rangle$ (i.e., the second and third
lowest Zeeman energy levels of Fig.\ \ref{Fig1boson}).
The full line
plots $m_{o}$, whereas the dash-dotted line plots
$2(m_{\parallel}-m_{\perp})$, both in units
of $\left(\pi\hbar^{2}a_{-}/m\right)$.  For reference, note
that the Ising form of the spin $\frac{1}{2}$ boson particle-particle
interaction takes the form
\begin{equation}
V\left(\vec{r}_{1},\vec{r}_{2}\right)=
\delta\left(\vec{x}_{1}-\vec{x}_{2}\right)
\left[ \lambda_{0}+m_{o} s_{1,z}s_{2,z}
+ 2 \left( m_{\parallel}-m_{\perp} \right) s_{1,z}s_{2,z}
\right],
\end{equation}
where $\lambda_{0}$ denotes the spin-independent
interaction strength in the Ising form,
$\lambda_{0}
= (\pi\hbar^{2} / m)\left[ a_{\uparrow}+a_{\downarrow}
+ 2 a_{u}\right]$.
Note that $m_{o}$ changes sign near
$b=0.535$, so that there exists an external magnetic field
strength at which the spin-independent effective short-range
magnetic field carried by the interacting particles can be
made to vanish (without having to take recourse to a Feshbach
resonance).  The Ising spin-spin interaction coefficient does
not change sign but it's magnitude is maximized at a
magnetic field $b\sim1.75$.  The precise values of the
magnetic fields at which the interaction parameters
exhibit this behavior may be different, but we expect the
DIS-approximation to give the correct qualitative behavior
even if the DIS-approximation is not expected to be accurate
at higher magnetic field values.

\section{A controllable $N$ boson quantum magnet}

The spin-spin forms of the effective inter-particle interactions reveal the analogy with
magnetic systems.  As an illustration we consider a specific system that promises a particularly
powerful and interesting simulation of a quantum magnet: N indistinguishable bosons
occupying two hyperfine spin states $\left\vert \uparrow\right\rangle $, $\left\vert \downarrow\right\rangle $,
confined by a tight spatial potential (which could be a single well of an optical lattice).
We assume that both $\left\vert \uparrow\right\rangle $ and $\left\vert \downarrow\right\rangle $ experience
the same trapping potential $V_{\uparrow}\left(\vec{x}\right)=V_{\downarrow}\left(\vec{x}\right)=
V_{T}\left(\vec{x}\right)$ of single-particle ground state $\chi_{T}\left(\vec{x}\right)$ and
single particle ground state energy $e_{T}$.  When $e_{T}$ exceeds all other
energy-per-particle values and the system relaxed to its motional ground state, all of the
$N$ bosons occupy the $\chi_{T}$-orbital and the spatial degrees of freedom are
`frozen', allowing only spin dynamics.  We also assume that the $\left\vert \uparrow\right\rangle $ and $\left\vert \downarrow\right\rangle $ experience a coherent two-photon Raman coupling which
can be effected by pulses of near-resonant lasers (or by means of an oscillating magnetic
field).  The resonant Raman coupling also introduces a detuning $\epsilon$ which acts as
an effective energy difference between $\left\vert \uparrow\right\rangle $ and $\left\vert \downarrow\right\rangle $ and the Raman coupling is described by a term
\begin{eqnarray}
& \sum_{j=1}^{N}\left[  -E_{R}\left(  t\right)  \left(  \left\vert
\downarrow\right\rangle _{j}\;_{j}\left\langle \uparrow\left\vert +\right\vert
\uparrow\right\rangle _{j}\;_{j}\left\langle \downarrow\right\vert \right)
+\frac{\epsilon}{2}\left(  \left\vert \uparrow\right\rangle _{j}%
\;_{j}\left\langle \uparrow\left\vert -\right\vert \downarrow\right\rangle
_{j}\;_{j}\left\langle \downarrow\right\vert \right)  \right]
\nonumber\\
& =-2E_{R}\left(  t\right)  \sum_{j=1}^{N}\;s_{j,x}\;+\epsilon\sum_{j=1}%
^{N}\;s_{j,z},
\end{eqnarray}
in the Hamiltonian. In the above expression, $E_{R}$ denotes the Rabi-coupling
energy which varies in time if the coupling is caused by a pulse.

The premise of a trapping potential sufficiently tight to freeze out the
spatial degrees of freedom translates into an $N$-particle wavefunction of the
type
\begin{equation}
\Psi\left(  \vec{r}_{1},\vec{r}_{2},....,\vec{r}_{N}\right)  =\chi_{T}\left(
\vec{x}_{1}\right)  \chi_{T}\left(  \vec{x}_{2}\right)  ...\chi_{T}\left(
\vec{x}_{N}\right)  \;\left\vert S^{N}\left(  \vec{s}_{1},\vec{s}_{2}%
,...\vec{s}_{N}\right)  \right\rangle ,
\end{equation}
where $\left\vert S^{N}\right\rangle $ denotes the spin state of the $N$-boson
system. By virtue of permutation symmetry -- the full wavefunction has to be
even under permutation of the full $\vec{r}=\left(  \vec{x},\vec{s}\right)  $
coordinates of any pair of particles -- the $\left\vert S^{N}\right\rangle $
spin state is required to be even with respect to the permutation of any pair
of spin variables $i$ and $j$. This condition limits the spin states to the
manifold of maximal spin magnitude. Specifically, if we introduce the total
spin operator, $\mathbf{S}^{N}=\sum_{j=1}^{N}\mathbf{s}_{j}$, then
$\mathbf{S}^{N}\cdot\mathbf{S}^{N}\left\vert S^{N}\right\rangle =\frac{N}%
{2}\left(  \frac{N}{2}+1\right)  \left\vert S^{N}\right\rangle $,
corresponding to a total spin magnitude $\frac{N}{2}$, i.e., all pseudo-spins
aligned. Hence, the spin state is a linear combination of $S^{N}=\frac{N}{2}$
states of total spin projection $M_{S}=-\frac{N}{2},...,\frac{N}{2}$,
$\left\vert S^{N}=\frac{N}{2},M_{S}\right\rangle $ with $S_{z}^{N}\left\vert
S^{N}=\frac{N}{2},M_{S}\right\rangle =M_{S}\left\vert S^{N}=\frac{N}{2}%
,M_{S}\right\rangle $. The $M_{S}=\frac{N}{2}$ is an $N$-spin stretched
state.
\begin{equation}
\left\vert S^{N}=\frac{N}{2},M_{S}=\frac{N}{2}\right\rangle =\left\vert
\uparrow\right\rangle _{1}\;\left\vert \uparrow\right\rangle _{2}%
\;...\;\left\vert \uparrow\right\rangle _{N},
\end{equation}

In deriving the expression for the total energy $E$ of the $N$ interacting boson system,
we use the Ising spin-spin form of the short-range particle-particle interaction.  Integrating
out the position variables explicitly, we encounter a volume $v$, the `trap volume' that
characterizes the `tightness' of the confining $V_{T}$,
\begin{equation}
\frac{1}{v} = \int d^{3} \vec{x} \; \vert \chi_{T}\left(\vec{x}\right) \vert^{4} \; ,
\end{equation}
For instance, the spin-independent interaction energy per particle, $e_{0}$, is
inversely proportional to the trap volume $v$ and proportional to the spin-independent interaction
strength in the Ising form, $\lambda_{0}=\left(4\pi\hbar^{2}/m\right) \left[ \frac{a_{\uparrow}+a_{\downarrow}
+ 2 a_{u}}{4}\right]$,
\begin{equation}
e_{0}=\frac{\lambda_{0}}{v}.
\end{equation}
The analogous spin-interaction energies per particle are given by the expressions
\begin{eqnarray}
\epsilon_{o} &  =\;\frac{m_{o}}{v}\;=\frac{4\pi\hbar^{2}}{m}\left(
\frac{a_{\uparrow}-a_{\downarrow}}{2}\right)  \frac{1}{v},\nonumber\\
\epsilon_{I} &  =\frac{2\left(  m_{\parallel}-m_{\perp}\right)  }{v}%
\;=\;\frac{4\pi\hbar^{2}}{m}\left(  a_{\uparrow}+a_{\downarrow}-2a_{u}\right)
\frac{1}{v}.
\end{eqnarray}
The total many-body energy, $E$, takes the form
\begin{eqnarray}
E &  =N\left[  e_{T}+\frac{\left(  N-1\right)  }{2}e_{0}\right]  +\left(
N-1\right)  \epsilon_{o}\sum_{j=1}^{N}\left\langle S^{N}\left\vert
s_{j,z}\right\vert S^{N}\right\rangle
\nonumber\\
&  +\frac{\epsilon_{I}}{2}\sum_{i\neq j}\left\langle S^{N}\left\vert
s_{j,z}s_{j,z}\right\vert S^{N}\right\rangle +\sum_{j=1}^{N}\left\langle
S^{N}\left\vert \left(  -2\epsilon_{R}s_{j,x}+\epsilon s_{j,z}\right)
\right\vert S^{N}\right\rangle ,
\end{eqnarray}
so that the integration over the position variable with short-range
interactions maps the $N$ spin-1/2 boson problem into that of $N$ 1/2 spins
coupled via an infinite range spin-spin interaction.  By adding and subtracting
\begin{equation}
\frac{\epsilon_{I}}{2}\sum_{j=1}^{N} s_{j,z}s_{j,z} = \frac{N \epsilon_{I}}{8},
\end{equation}
we cast the Hamiltonian in terms of the total spin-operator $\mathbf{S}^{N}=
\sum_{j=1}^{N}\mathbf{s}_{j}$.  We also define effective magnetic fields that are
$c$ numbers,
\begin{eqnarray}
\mathbf{H}_{o}=\left( N -1 \right) \mathbf{z} \epsilon_{o}
\nonumber \\
\mathbf{H}_{R}=-2 \epsilon_{R} \mathbf{x}+ \epsilon \mathbf{z}
\end{eqnarray}
The energy $E$, up to an unimportant shift,
$E^{\prime}=E-N\left[
e_{T}+\left(  N-1\right)  \frac{e_{0}}{2}\right]  -\frac{N\epsilon_{I}}{8}$, then takes the
form
\begin{equation}
E^{\prime}=\left\langle S^{N}\left\vert \mathbf{S}^{N}\cdot\left(
\mathbf{H}_{o}+\mathbf{H}_{R}\right)  +\frac{\epsilon}{2}S_{z}^{N}S_{z}%
^{N}\right\vert S^{N}\right\rangle ,
\label{spinh}
\end{equation}
reminiscent of the Hamiltonian of magnetic single domain grains with anisotropic
spin-spin interactions \cite{gun}.  Chudnovsky and Gunther had pointed out that the anisotropy
can set conditions under which we expect macroscopic quantum tunneling: sufficiently
strong exchange interactions force the individual spins to align into a macroscopic
spin vector, the anisotropy can give local energy minima corresponding to two distinct
directions of the macroscopic spin and quantum mechanically, the macroscopic spin
can travel through a classically forbidden region giving tunneling although the expected
rate for such processes are exponentially suppressed with the number of spins.
In the $N$-boson quantum magnet, the alignment is enforced by permutation symmetry,
the anisotropy caused by the Ising (or XY) nature of the effective inter-particle interactions
and the number of bosons can, in principle, be controlled experimentally.

We recognize the operator in the spin bracket of Eq.\ (\ref{spinh}) as the spin
Hamiltonian $\hat{H}$.  The Heisenberg equation of motion for the total spin operator,
\begin{equation}
i \hbar \frac{ d\mathbf{S}^{N}}{dt}
= \left[ \mathbf{S}^{N}, \hat{H} \right]_{-}
= i \mathbf{S}^{N} \times \left[ \mathbf{H}_{o}+\mathbf{H}_{R} \right]
+ i \epsilon_{I} \mathbf{S}^{N}_{\parallel} \left( \mathbf{S}^{N} \cdot \mathbf{z} \right) \; ,
\end{equation}
where $\mathbf{S}^{N}_{\parallel}$ denotes the part of the total spin vector that points in
the direction of the magnetic field, $\mathbf{S}^{N}_{\parallel} = \mathbf{z} \; \left(\mathbf{S}^{N}
\cdot \mathbf{z}\right)$, yields an Ehrenfest type of equation for the total spin expectation value
if we take its expectation value.
We can write the resulting equation as a Landau-Lifshitz equation (without damping term)
\cite{landau},
\begin{equation}
\frac{d}{dt} \langle \mathbf{S^{N}} \rangle
= \frac{1}{\hbar} \langle \mathbf{S}^{N} \times \mathbf{H}_{total} \rangle \; \; ,
\end{equation}
where the total magnetic field $\mathbf{H}_{total}$ includes the Raman coupling
effective field, the the effective, spin-independent short-range field carried
by the other particles and the contribution caused by the Ising spin-spin interactions,
\begin{equation}
\mathbf{H}_{total} = \mathbf{H}_{o}+\mathbf{H}_{R} + \epsilon_{I} \mathbf{S}^{N}_{\parallel}
\left( \mathbf{S}_{N} \cdot \mathbf{z} \right) \; .
\end{equation}
In the absence of Raman coupling $\mathbf{H}_{R}=0$ the total magnetic field $\mathbf{H}_{total}$
points in the z-direction and all the spin expectation vector can do is precess around the z-direction.
In fact, the derivative of the expectation value of any power of $S^{N}_{z}$ vanishes so that
the conservation of up and down particles ensures that the distribution of the spin up and
spin down particles will remain constant in time precluding collective tunneling of the spin
in the absence of Raman coupling.   A Raman pulse can then precisely control and initiate the
macroscopic quantum tunneling while leaving the other assumptions and parameters of the
system untouched.  In addition to tunneling, the ground state of the system can be
a superposition of two distinct states in each of which the total spin points in different
directions.  That Raman-coupled (or Josephson-coupled) two-component BEC systems
can take on macroscopic Schrodinger cat states was pointed out in \cite{Cirac} and
worked out in \cite{Gordon} -- we simply determine the interaction parameters and indicate
how these can be controlled.  The Raman control provides an important advantage to the
N-boson quantum magnet over the double-well proposals for realizing macroscopic quantum
tunneling and creating macrocopic Schrodinger cat states.  Varying the potential barrier in
a double well system to control the tunneling can also render the two-state approximation invalid
and lead to unwanted excitations.

Observing the coherent oscillations of the total spin that is quantum tunneling can also test
fundamental aspects of quantum mechanics (against macroscopic realism) by verifying
Legget-Garg inequalities \cite{legg2}.  In addition, the $N$-boson quantum magnet spin dynamics can also explore
spin squeezing, non-classical quantum evolution near unstable trajectories \cite{vardi} and, when the
Ising interaction is eliminated by a Feshbach resonance, realize the Burnett-Holand proposal for
Heisenberg limited interferometry \cite{Burnett} by using the Raman pulse as a
beam-splitter.  These connections become obvious using the spin-spin form of the inter-particle
interactions which also reveal the control that routine cold atom knobs such as the
intensity of the confining potential, the detuning of the Raman coupling pulse and
the magnetic field of a Feshbach resonance can exercise.

\section{Conclusions}

In conclusion, we described the effective spin-dependent interactions
of ultra-cold alkali atoms occupying two distinct hyperfine states in
an external magnetic field.  The magnetic field lifts the degeneracy of
the atomic Zeeman levels and permits the selection of two hyperfine
states to act as the effective `spin-up' and `spin-down' states of the
particles so the atoms can mimic the behavior of magnetic spin-$\frac{1}{2}$ particles.
We described the spin-dependent effective interaction as a spin-spin
interaction.  The form of the effective spin-spin interaction depends explicitly on the
quantum statistics of the interacting particles.  As a consequence of the
zero-range nature of the interaction, the interaction of spin-$\frac{1}{2}$ bosons
can be described as an Ising or, alternatively, as an $XY$-coupling.
The parameters of the spin-spin interaction depend on the
scattering lengths of the relevant binary alkali atom collision channels
in the external magnetic field.  For relatively low values of the magnetic
field (sufficiently large to cause a Zeeman level splitting that permits
the selection of two hyperfine levels) we calculated the parameters as
a function of the external magnetic field in the Degenerate Internal
State (DIS) approximation.  We illustrated the advantage of the spin-spin
interaction form by mapping the system of N spin-$\frac{1}{2}$ bosons in a tight
trapping potential on that of N spin-$\frac{1}{2}$ spins coupled via an infinite
range interaction.  The explicit expressions reveal which parameters of
the spin Hamiltonian can be controlled and how.  The spin Hamiltonian
also suggests that the N-boson quantum magnet provides an intriguing
laboratory for the exploration of fundamental quantum studies.
The list of promising uses include the study of collective quantum spin
tunneling (which can be used for testing fundamental aspects of quantum
mechanics), the controlled observation and utilization of spin squeezing
and the creation and study of highly non-classical states.

\section{Acknowledgments}

The work of one of the authors, E.T., was supported by the Los Alamos
Laboratory Directed Research and Development (LDRD) program.

\end{document}